\documentclass[%
reprint,
 amsmath,amssymb,
 aps,
prb,
]{revtex4-2}

\usepackage{graphicx}
\usepackage{dcolumn}
\usepackage{bm}
\usepackage{lipsum,mwe}
\usepackage{amsthm,amsmath,amssymb}
\usepackage{mathrsfs}
\usepackage{subfigure}
\usepackage{hyperref} 
\hypersetup{hypertex=ture,colorlinks=true,linkcolor=blue,anchorcolor=blue,citecolor=blue}
\usepackage{caption}

\usepackage{setspace}
\fontsize{20pt}{24pt}
\usepackage{mathtools}

\begin{document}

\preprint{APS/123-QED}

\title{Fragile topological phase on the triangular kagome lattice and its bulk-boundary correspondence}

\author{Yun-Feng Chen}
\author{Dao-Xin Yao}%
\email{yaodaox@mail.sysu.edu.cn}
\affiliation{%
State Key Laboratory of Optoelectronic Materials and Technologies, School of Physics, Sun Yat-Sen University, Guangzhou 510275, China
}%

\date{\today}

\begin{abstract}
We predict and examine various topological states on a two-dimensional (2D) triangular kagome lattice (TKL) using the tight-binding (TB) models and theory of topological quantum chemistry (TQC). Firstly, on the basis of TQC, we diagnose band structures with fragile topology and calculate Wilson-loop spectra and Hofstadter butterfly spectra to confirm their non-trivial nature.
Secondly, we examine the bulk-boundary correspondence and find that an obstructed-atomic-limit (OAL) insulator hosts fractional corner states without being accompanied by fragile topological band structures, which implies that the presence of OALs and corner states is not a sufficient condition to fragile topology. Last but not least, we predict a topological phase transition from a second-order topological phase to a first-order topological phase that can be realized in the TKL under the action of a magnetic field.
\end{abstract}

\maketitle


\begin{figure*}[htbp]
	\centering
	\subfigure[]{
		\includegraphics[width=5.5in]{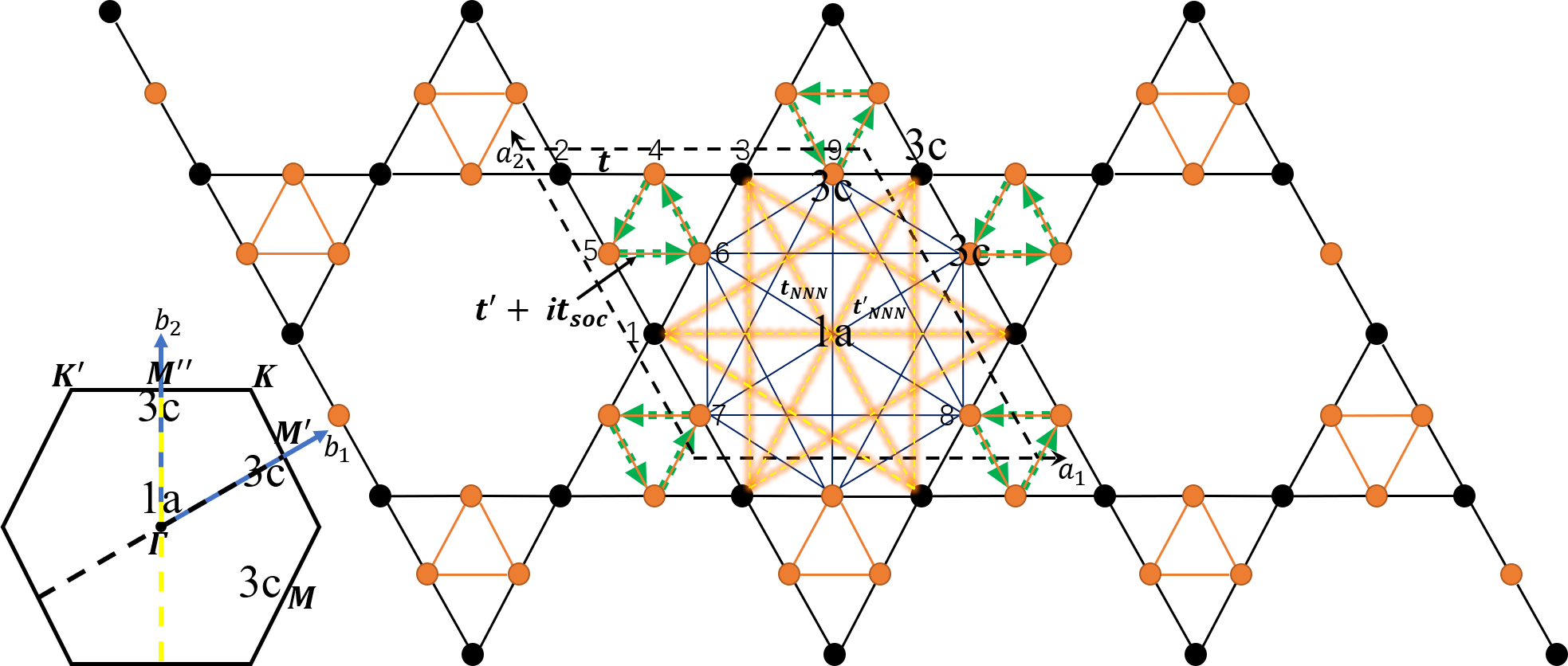}
		\label{fig.1a}
	}
	\\
	\subfigure[]{
		\includegraphics[width=1.95in]{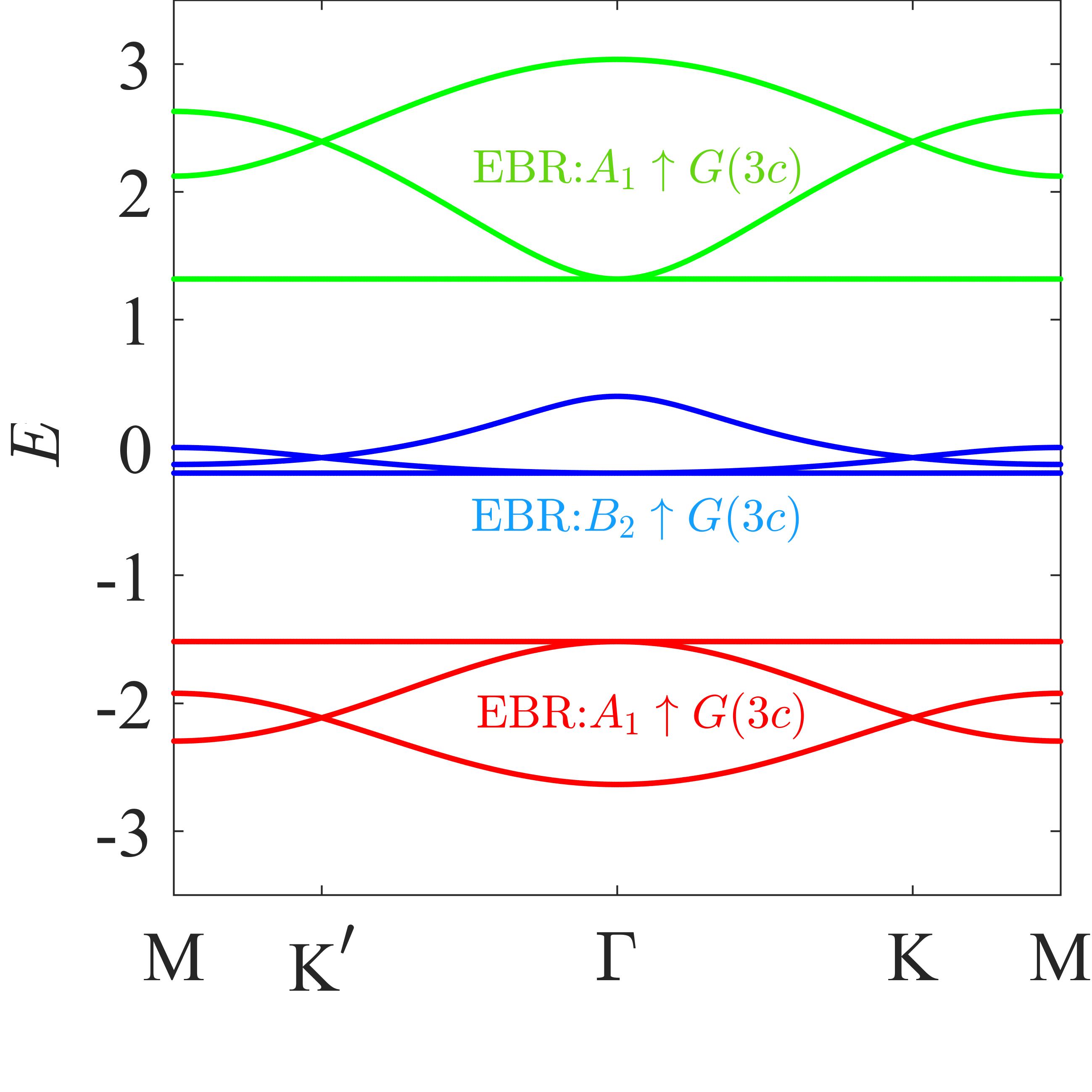}
		\label{fig.1b}
	}
	\subfigure[]{
		\includegraphics[width=2in]{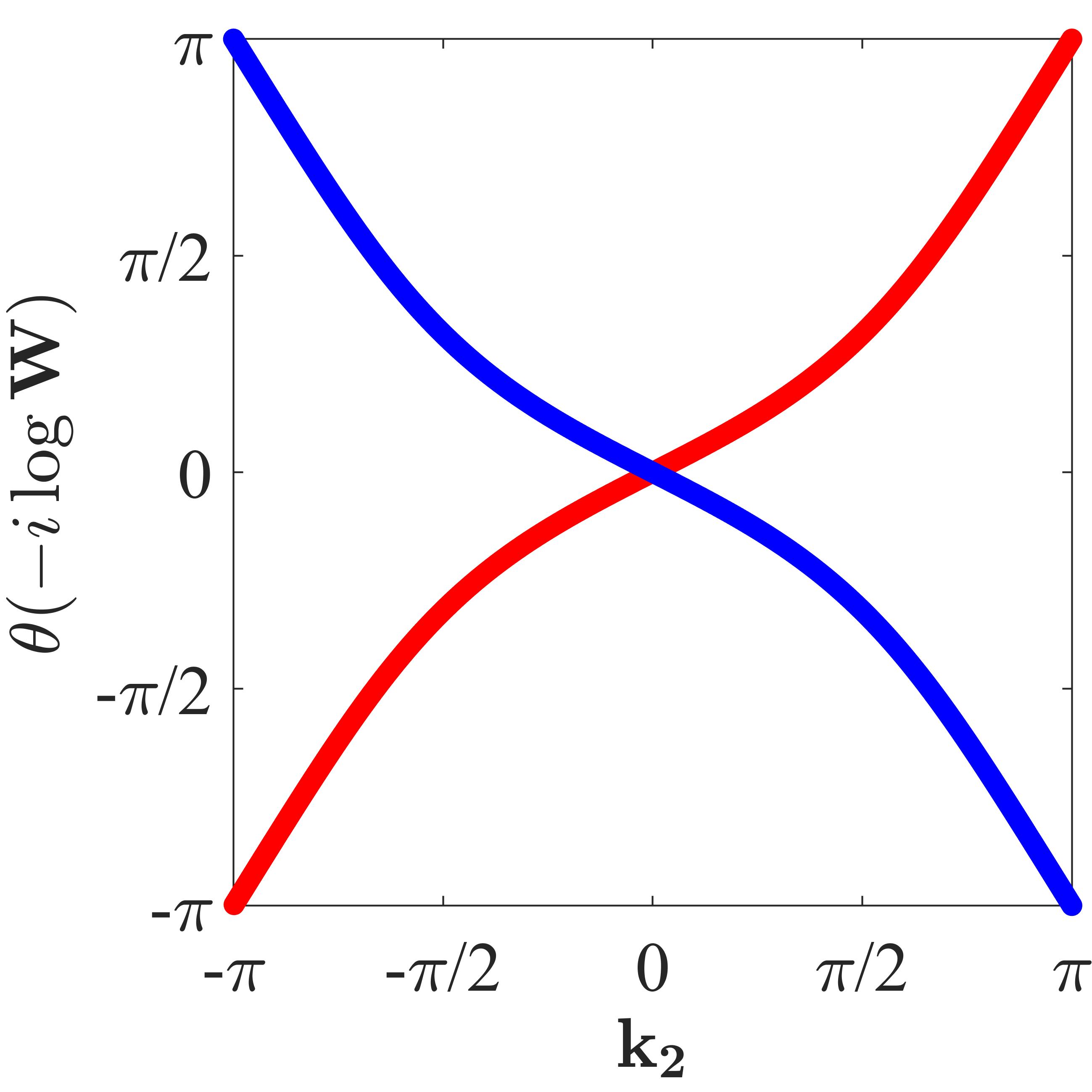}
		\label{fig.1c}		
	}
	\subfigure[]{
		\includegraphics[width=2in]{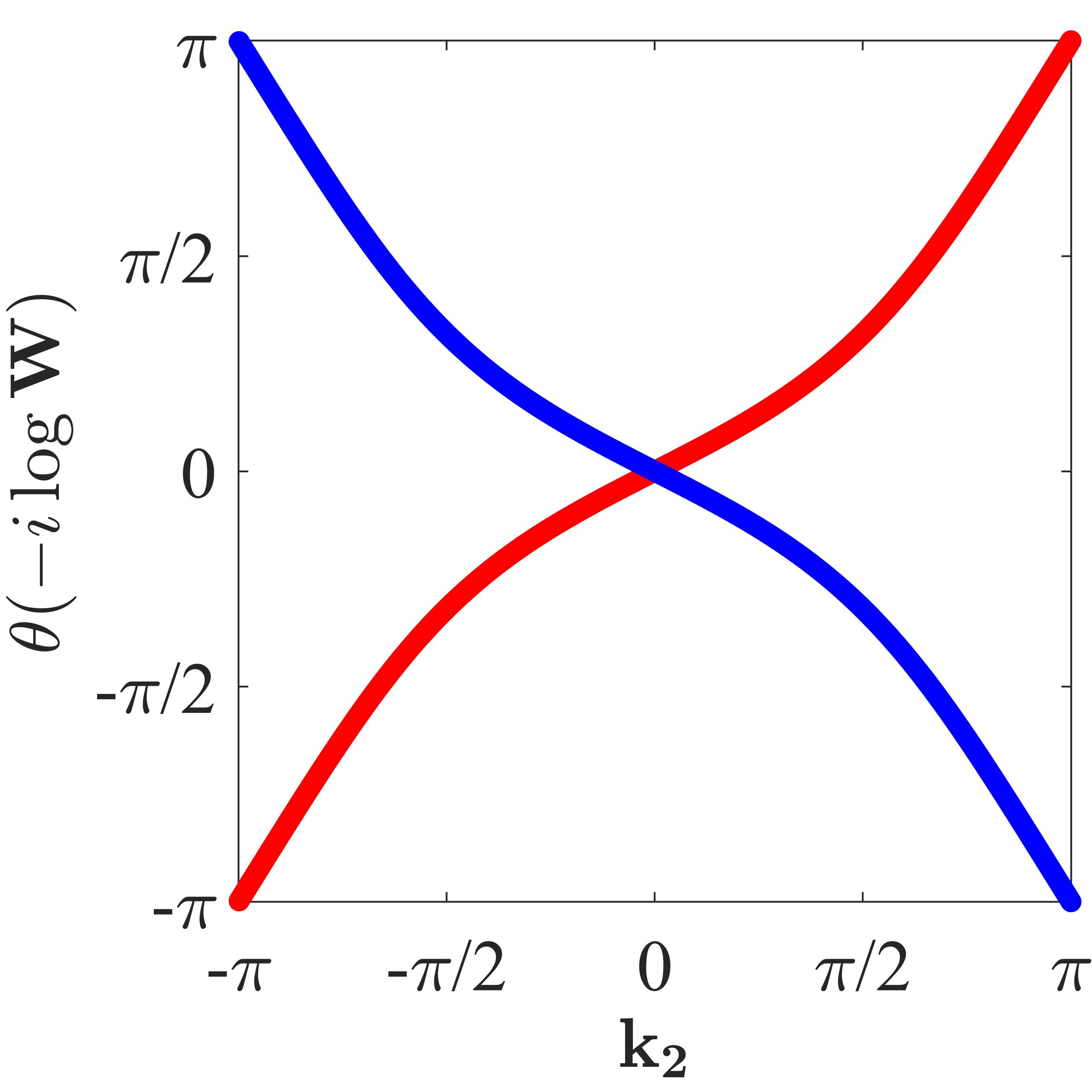}
		\label{fig.1d}
	}
	\caption{(a) Diagram of a TKL and its first BZ. $\boldsymbol{a_1}$, $\boldsymbol{a_2}$, $\boldsymbol{b_1}$, $\boldsymbol{b_2}$ are basis vectors in real space and momentum space respectively. NN hoppings with energies $t$ and $t^{\prime}$ are represented by black and orange bonds, respectively. NNN hoppings with energies $t_{_{\emph{NNN}}}$ and $t_{_{\emph{NNN}}}^{\prime}$ are represented by yellow dashed and blue solid bonds, respectively. Those anticlockwise Haldane-model-like hoppings with energy $t_{soc}$ are represented by green dashed arrows. Wyckoff positions 1a and 3c are indicated in this diagram. (b) Band structure of $H_{_{\emph{NN}}}^{^{\emph{TB}}}$ along the path $M$-$K^{\prime}$-$\Gamma$-$K$-$M$ (in $\boldsymbol{b_1}$+$\boldsymbol{b_2}$ direction), with $t=1$ and $t^{\prime}=0.2t$. The top, middle, and bottom groups and their corresponding EBRs are indicated in this spectrum.
	$\left[(c),(d)\right]$ Wilson-loop spectra of the upper and lower sets of FTBs. Red and blue spectral lines belong to different FTBs from the same set. Despite that the upper set is flatter than the lower set, they exhibit the same Wilson-loop spectrum because of their identical momentum-space representations. The Wilson loop along $\boldsymbol{b_2}$ and the integral loop along $\boldsymbol{b_1}$ are indicated by yellow and black dashed lines in BZ, respectively.}
	\label{fig.1}
	
\end{figure*}

\section{\label{sec:level1}Introduction}\label{section1}
Since the discovery of integer quantum Hall effect, robust bulk-boundary correspondence has become a hallmark of non-trivial topology. Later, the discovery of high-order topological insulators (HOTIs) enriches the physics with robust high-order bulk-boundary correspondence (e.g., bulk-hinge correspondence and bulk-corner correspondence) \cite{schindler2018higher,schindler2018higher2,xu2019higher,kunst2018lattice,xue2019realization,rodriguez2019higher}. Namely, a (N-n)-order topological insulator is a N-dimensional insulator which hosts topological states on its n-dimensional boundaries. Recently, however, this seemly complete scheme is challenged by the fact that fragile topological phases (FTPs) violate the bulk-boundary correspondence and only host chiral edge states under some specific twisted boundary conditions \cite{zou2018band,ahn2019failure,song2021twisted,lian2020landau,wu2021landau,chou2020hofstadter,lu2021multiple,lu2020fingerprints,bradlyn2017topological,bradlyn2019disconnected,cano2018topology,cano2018building,po2018fragile,bouhon2019wilson,bouhon2020geometric,peri2021fragile,song2020twisted,de2019engineering,zhang2021tunable,peri2020experimental,yu2021dynamical,kruthoff2017topological,slager2013space,chiu2020fragile,ma2020spin}. Besides, those widely-used topological invariants that are proposed to diagnose stable non-trivial topology (e.g., Chern number, Kane-Mele invariant) fail to capture the FTPs because of the fragility of these phases manifested by the addition of trivial bands \cite{bradlyn2017topological,bradlyn2019disconnected,cano2018topology,cano2018building,po2018fragile,bouhon2019wilson}. More specifically, the Wannier obstructions of the FTPs (i.e., the topological obstacles that prevent FTPs from transforming into atomic limits which are described by exponentially localized Wannier functions) can be resolved only by coupling these phases with trivial bands, which is totally outside the K-theory framework. In this situation, the theory of topological quantum chemistry (TQC) has been developed to characterize the fragile topology of FTPs \cite{bradlyn2017topological,bradlyn2019disconnected,cano2018topology,cano2018building,po2018fragile,bouhon2019wilson}. Its central idea is to construct mappings from real-space orbitals to momentum-space topology. According to this theory, a set of exponentially localized, decoupled orbitals on maximal Wyckoff positions (the high-symmetry positions of a space group) in the real space can induce a non-local space-group representation in the momentum space, which is the so-called band representation (BR) or elementary band representation (EBR) \cite{bradlyn2017topological,bradlyn2019disconnected,cano2018topology,cano2018building,po2018fragile,bouhon2019wilson}. Mathematically, a BR is a direct sum of EBRs. A band structure shares the triviality of an atomic limit if its irreducible representations of the point groups at the momenta in the first Brillouin zone (BZ) are the same as those of a band representation; otherwise, this band structure carries non-trivial topology \cite{bradlyn2017topological}. More specifically, a subtraction of EBRs, which topologically can not be transformed into a BR, maps to a set of fragile topological bands (FTBs), and a representation that falls outside the trivial and fragile-topological classes maps to a band structure with stable topology \cite{bradlyn2017topological}. The former mapping properly manifests the fragility of fragile topology: the obstruction of transforming a subtraction of EBRs into a BR disappears after the addition of those EBRs that have been subtracted. In fact, the subtraction of EBRs represents a split band structure that consists of two subspaces: in most cases, the lower one is fragile topological and the upper one possesses a ``practical'' Wannier obstruction \cite{luo2021fragile,shang2020second,PhysRevLett.123.186401,lange2021subdimensional,herzog2020hofstadter}. Namely, electronic bands with this obstruction adiabatically connect to the so-called obstructed atomic limits (OALs) \cite{bradlyn2017topological} where the localized orbitals do not coincide with any of the ionic sites in the unit cell. These bands are hereafter referred to as OALs for simplicity. The presence of an OAL causes the mismatch between the locations of ionic sites and the centers of electrons, which eventually gives rise to extra fractional charge distribution on the corners of a finite lattice. Therefore, when a FTP is complementary to an OAL (i.e., all the bands under the OAL, including the FTBs, are filled), it hosts fractional corner states under open boundary conditions. \cite{luo2021fragile,shang2020second,PhysRevLett.123.186401,lange2021subdimensional,herzog2020hofstadter}.

Although stable topological invariants fail to distinguish FTPs from normal trivial phases, there are clear differences between them. For example, FTPs react differently towards a perpendicular magnetic field. In this magnetic field, electrons on 2D lattices exhibit a kind of fractal structures in energy spectra, called the Hofstadter butterly \cite{hofstadter1976energy}. Since the landau levels (LLs) of a set of trivial electronic bands are bounded inside the energy span of this set in the zero-field limit, the Hofstadter butterfly of a trivial set shall never connect to the one of another trivial set, if these sets are isolated from each other before the presence of the magnetic field. Interestingly, FTPs exhibit connected Hofstadter butterfly spectra, which manifests their non-trivial nature \cite{lian2020landau,wu2021landau,chou2020hofstadter,lu2021multiple,lu2020fingerprints,herzog2020hofstadter,rhim2020quantum,hwang2021geometric,guan2021landau}. Moreover, FTPs also exhibit non-trivial windings of Berry phases in Wilson-loop spectra, which definitely indicates their non-trivial topology. Therefore, these spectra have been widely analyzed to study the difference between FTPs and normal trivial phases. In this work, we focus on examining the FTP on a triangular kagome lattice (TKL) \cite{wang2018coexistence,yao2008x,chen2012kondo,loh2008dimers,loh2008thermodynamics,kim2017boron,gonzalez1993structural,maruti1994magnetic,mekata1998magnetic} and its bulk boundary correspondence. Compared to the kagome lattice, we find that the more frustrated structure of the TKL becomes an advantage for hosting various topological states since more complicated Wannier obstructions can be created. Last but not least, we also find a topological phase transition from a second-order topological phase (SOTP) to a first-order topological phase (FOTP) that can be realized in this lattice.

The rest of this paper is organized as follows: In Sec.\ref{section2a}, we use the theory of TQC to diagnose the band topology of several tight-binding models on a TKL. We compute the Wilson-loop spectra of two sets of FTBs in Sec.\ref{section2b} and examine the bulk-boundary correspondence in Sec.\ref{section2c}. In Sec.\ref{section3a}, we compute the Hofstadter butterfly spectra and discuss their bounded and connected patterns. In Sec.\ref{section3b}, we demonstrate how to realize the transition from a SOTP to a FOTP by increasing the magnetic field strength. Finally, we conclude in Sec.\ref{section4}.

\begin{figure*}[htbp]
	\centering
	\subfigure[]{
		\includegraphics[width=2in]{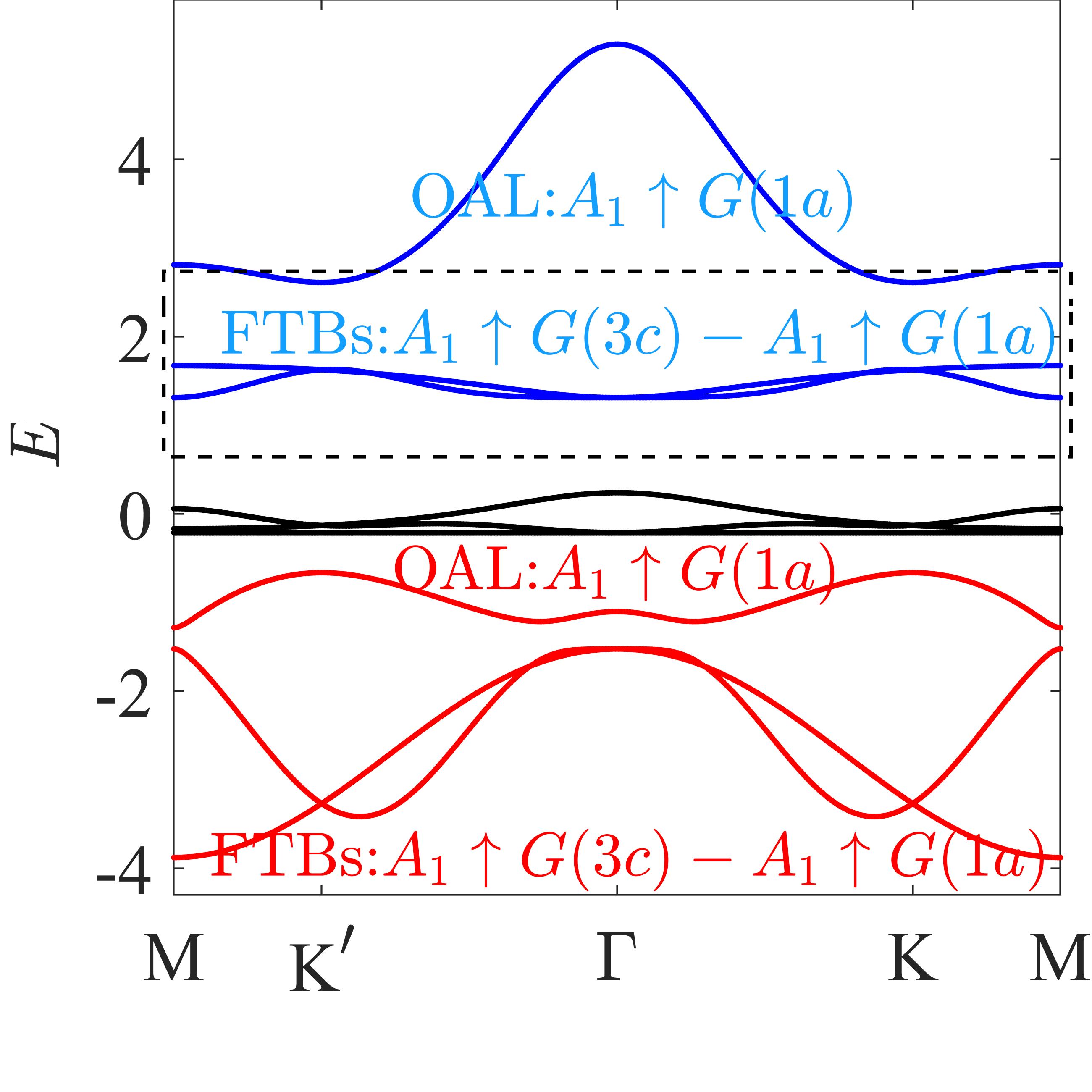}
		\label{fig2a}
	}
	\subfigure[]{
		\includegraphics[width=2in]{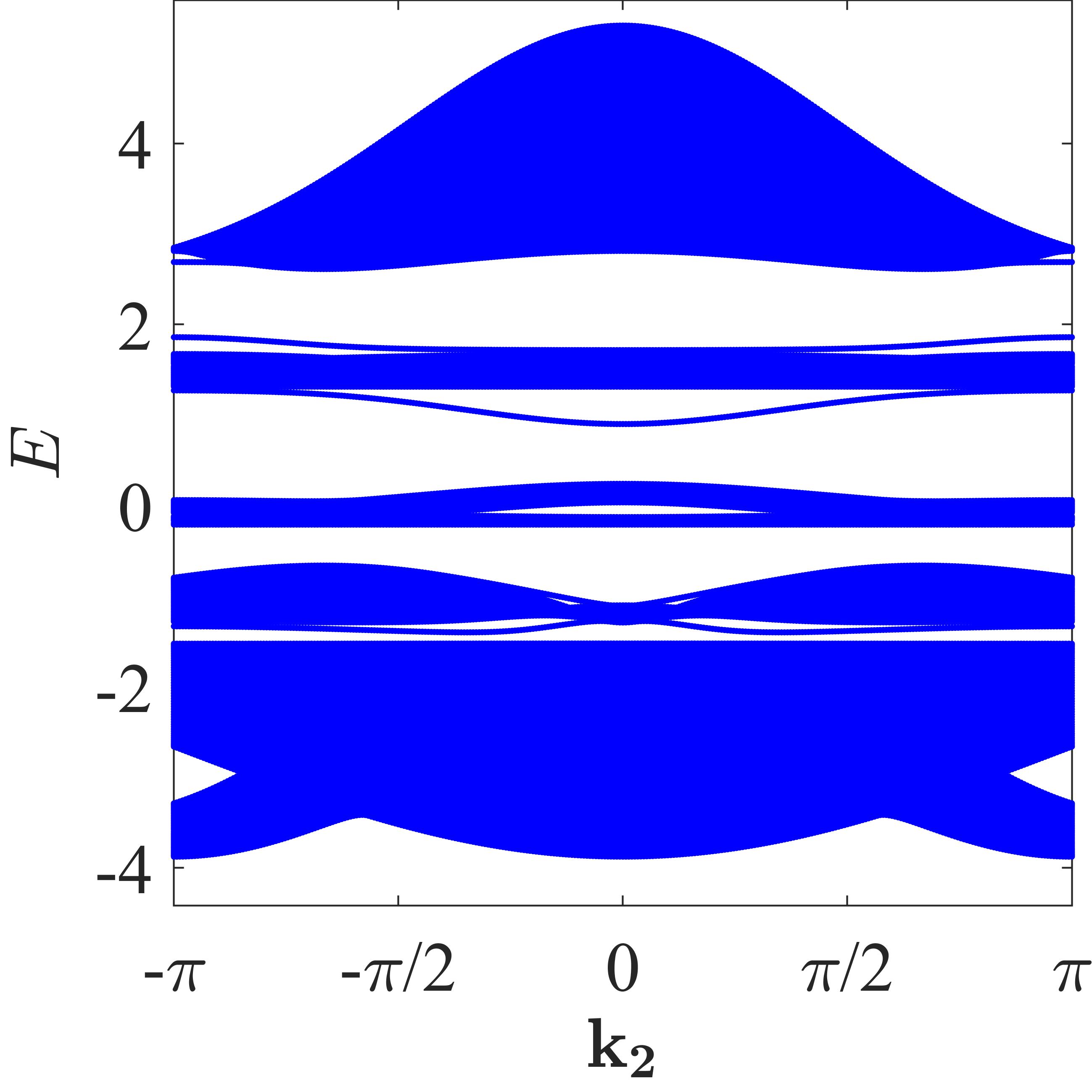}
		\label{fig2b}
	}
	\subfigure[]{
		\includegraphics[width=2.5in]{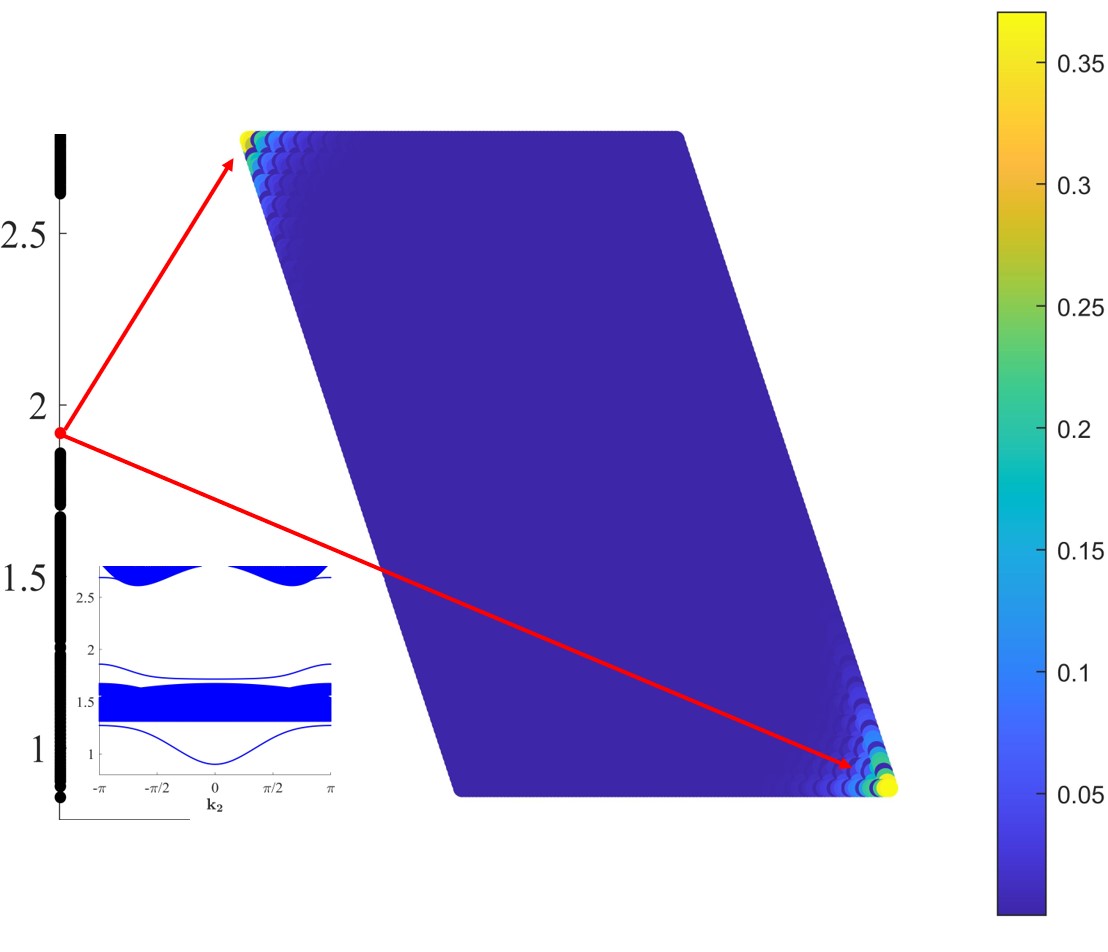}
		\label{fig2c}
	}
	\\
	\subfigure[]{
		\includegraphics[width=2in]{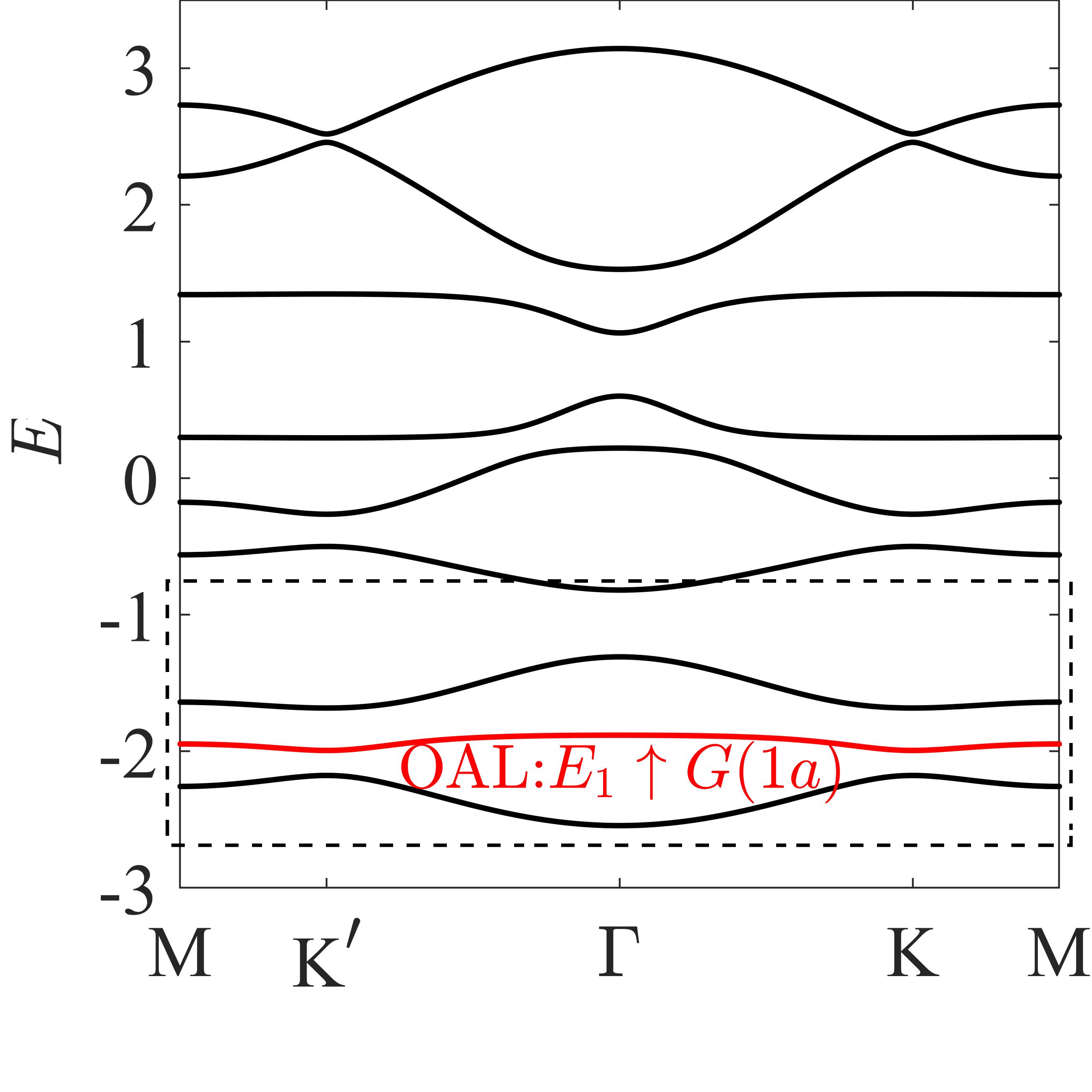}
		\label{fig2d}
	}
	\subfigure[]{
		\includegraphics[width=2in]{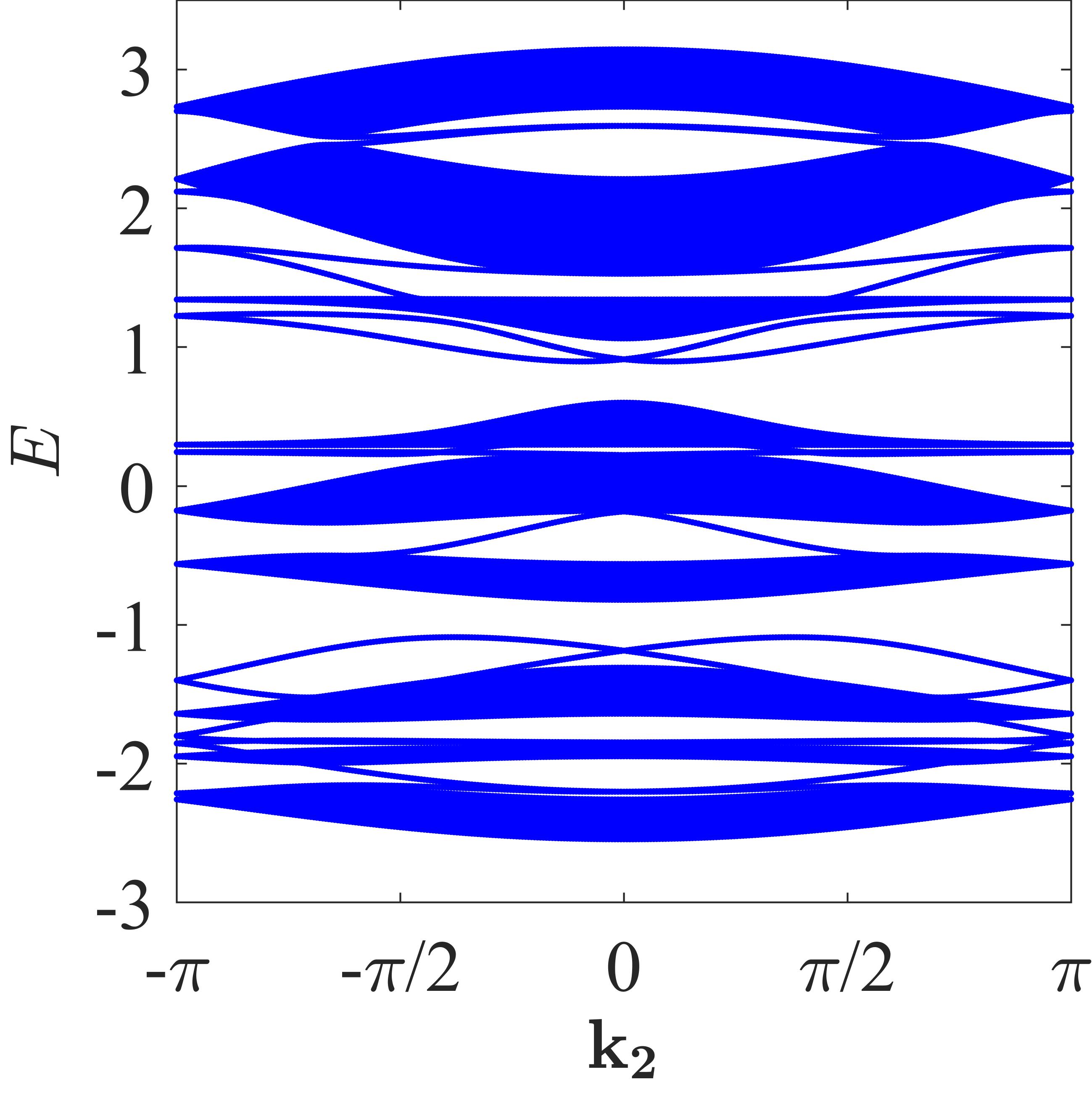}
		\label{fig2e}
	}
	\subfigure[]{
		\includegraphics[width=2.5in]{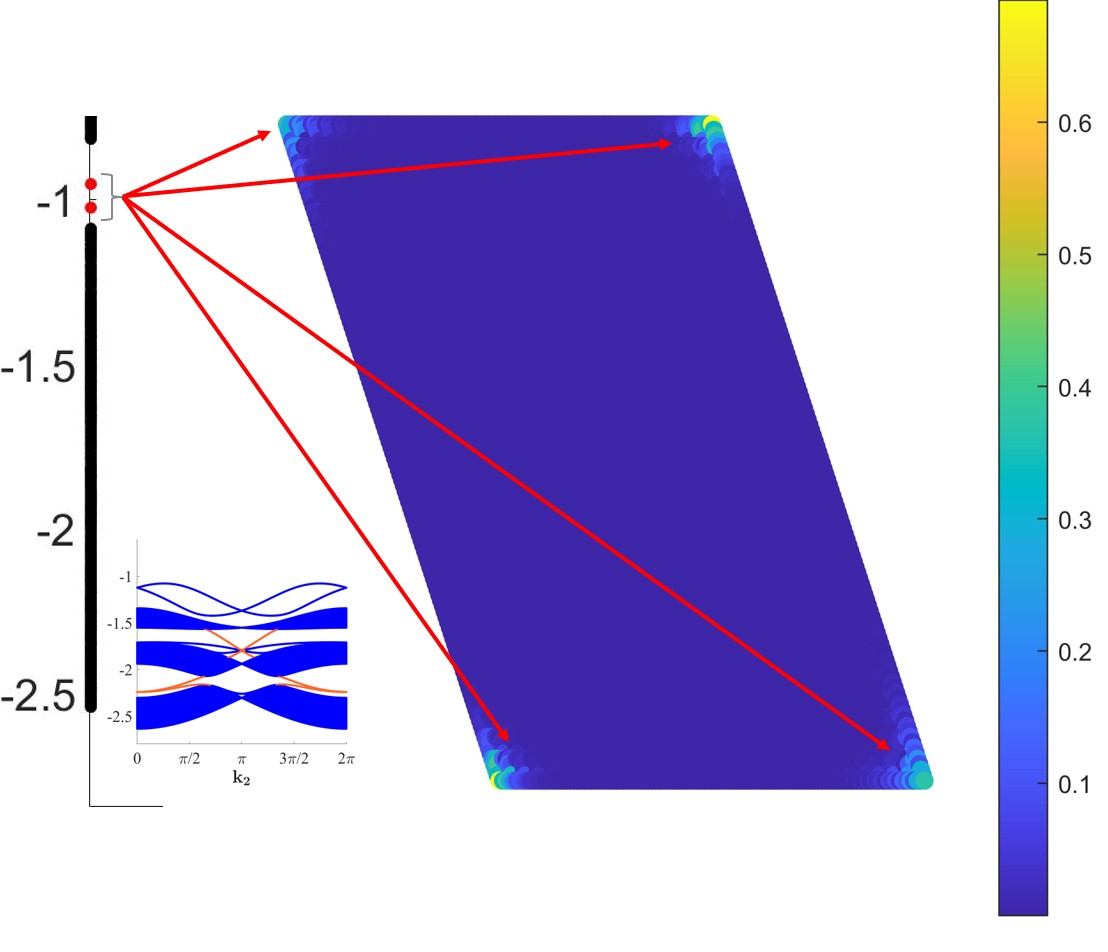}
		\label{fig2f}
	}
	
	\caption{$\left[\left(a\right)\sim\left(c\right)\right]$ Band structures of $H^{^{\emph{TB}}}_{_{\emph{NNN}}}$ for (a) periodic boundary conditions, (b) open boundary condition in $\boldsymbol{a_1}$ direction ($L_1=200$), and (c) open boundary conditions in $\boldsymbol{a_1}$ and $\boldsymbol{a_2}$ directions ($L_1=L_2=30$), with $t_{_{\emph{NNN}}}=0.6t$, $t_{_{\emph{NNN}}}^{\prime}=0.4t^{\prime}$, and $t^{\prime}$=$0.2t$. $\left[\left(d\right)\sim\left(f\right)\right]$ Similar band structures of $H^{^{\emph{TB}}}_{_{\emph{SOC}}}$ with $t_{soc}=0.3t$ and $t^{\prime}$=$0.3t$. 
    In (c) and (f), the energy spectrum on the left shows the fractional corner states marked by red dots, and the diagram on the right shows the distribution of the corner states on the 30$\times$30 TKL. Chiral edge states appear in (e), and those in the first and second gaps (counted from the bottom) are redrawed and highlighted in red in the insert in (f). Black dashed rectangles in (a) and (d) draw the energy ranges of the spectra and inserts in (c) and (f), respectively.}
	\label{fig2}
\end{figure*}

\section{Fragile topological bands and bulk-boundary correspondence}\label{section2}

\subsection{Symmetry analysis based on TQC}\label{section2a}
Firstly, we consider a spinless nearest-neighbor (NN) tight-binding (TB) model on the TKL [see Fig.\ref{fig.1a}]:
\begin{equation}
H_{_{\emph{NN}}}^{^{\emph{TB}}}=\sum_{<i,j>}\left[t\sum_{\alpha,\beta} C_{i,\alpha}^{\dagger}C_{j,\beta}+ t^{\prime}\sum_{\beta^{\prime},\beta}C_{i,\beta^{\prime}}^{\dagger}C_{j,\beta}\right]+H.c. ,
\end{equation}
where $\left\langle\ \right\rangle$ denotes that only the hoppings between nearest sites are considered in this summation. The lattice sites of the small and big triangles in the unit cell are labeled by $\alpha (\alpha^{\prime})$=1,2,3 and $\beta(\beta^{\prime})$=4,5,6,7,8,9, respectively. The space group of the TKL is P6mm (No.183), and the point groups of the three high-symmetry momenta in BZ (i.e., $\Gamma$=(0,0),K=$\left(\frac{2\pi}{3},\frac{2\pi}{3}\right)$, M=$\left(\frac{\pi}{2},0\right)$) are $C_{6v}$, $C_{3v}$ and $C_{2v}$, respectively. To be representative, we assume that the energies of the hoppings along the red and black bonds in Fig.\ref{fig.1a} is $t=1$ and $t^{\prime}=0.2t$, respectively. The resulting band structure is shown in Fig.1(b), where the nine bands are divided into the top, middle and bottom groups. By calculating the irreducible representations (see Table.\ref{table1}), we determine that the top and bottom groups are represented by the EBR $A_1\uparrow G(3c)$ and the middle group is represented by $B_2\uparrow G(3c)$. These two EBRs are induced by orbitals localized on the 3c Wyckoff positions [see Fig.\ref{fig.1a}] in the unit cell, where ionic sites coincide with the centers of the orbitals. Therefore, no OAL is created and these groups of bands are normally trivial. Next, We introduce the next-nearest-neighbor (NNN) hoppings shown in Fig.\ref{fig.1a} into the TB model to creat fragile topological band structures:

\begin{equation}
	\begin{aligned}
		&	 H_{_{\emph{NNN}}}^{^{\emph{TB}}}=H_{_{\emph{NN}}}+H_{_{\emph{NNN}}},\\
		&	
		H_{_{\emph{NNN}}}=\sum_{\mathclap{\left\langle\left\langle i,j\right\rangle\right\rangle}}\left[t_{_{\emph{NNN}}}\sum_{\alpha,\alpha^{\prime}} C_{i,\alpha}^{\dagger}C_{j,\alpha^{\prime}}+ t_{_{\emph{NNN}}}^{\prime}\sum_{\beta^{\prime},\beta}C_{i,\beta^{\prime}}^{\dagger}C_{j,\beta}\right]+H.c.,
	\end{aligned}
\end{equation}
where $\left\langle\left\langle\ \right\rangle\right\rangle$ denotes that the hoppings between next-nearest sites are also considered in this summation. With the parameters of the NN hoppings unchanged, we set the parameters of the NNN hoppings between $\alpha$ sites and between $\beta$ sites to be $t_{_{\emph{NNN}}}=0.6t$ and $t_{_{\emph{NNN}}}^{\prime}=0.4t^{\prime}$, respectively. The resulting band structure is shown in Fig.\ref{fig2a}, where the top and bottom groups are both split into an OAL and a set of FTBs (i.e., a two-band subspace hosting fragile topology), and the middle group remain unchanged, still represented by $B_2\uparrow G(3c)$. The two sets of FTBs, called the upper and lower sets, are represented by the same subtraction of EBRs (see Table.\ref{table1}): $A_1\uparrow G(3c)\ominus A_1\uparrow G(1a)$. It indicates that they are indeed fragile topological, which can be confirmed by calculating the Wilson-loop spectrum (see Sec.\ref{section2b}). As for the band represented by $A_1\uparrow G(1a)$, it is identified as an OAL since no ionic site coincides with the 1a Wyckoff position in the unit cell [see Fig.\ref{fig.1a}]. Even though the band structure seems to be completely changed, all the degeneracies at high-symmetry points are still preserved because the space-group symmetry remains unchanged. Besides, the NNN hoppings has no effect on time-reversal symmetry $\mathcal{T}$ because of their real hopping parameters.

	
\begin{table*}
	\caption{ The irreducible representations at $\Gamma$, $K$, and $M$ for the three groups of bands, the two sets of FTBs, and the three EBRs of P6mm. By comparing these irreducible representations, we find that the top and bottom groups matches the EBR $A_1\uparrow G(3c)$ and the middle one matches $B_2\uparrow G(3c)$. Similarly, we subtract $A_1\uparrow G(1a)$ from $A_1\uparrow G(3c)$ and obtain the representations of the two sets of FTBs. The notations used here are from the Bilbao Crystallographic Server \cite{aroyo2011crystallography,aroyo2006bilbao,aroyo2006bilbao2}. The number in brackets behind an irreducible representation denotes its dimensions, and the letter in brackets behind an EBR denotes the Wyckoff positions wherefrom this EBR is induced.}
	\label{table1}
	\begin{ruledtabular}
		\begin{tabular}{ccccc}
			
			Trivial bands&$\Gamma$&$K$&$M$\\ \hline
			top group&$\Gamma_1(1)\oplus\Gamma_5(2)$ &$K_1(1)\oplus K_3(2)$ &$M_1(1)\oplus M_3(1)\oplus M_4(1)$
			\\
			middle group&$\Gamma_4(1)\oplus\Gamma_6(2)$
			&$K_2(1)\oplus K_3(2)$&$M_1(1)\oplus M_2(1)\oplus M_4(1)$
			\\
			bottom group&$\Gamma_1(1)\oplus\Gamma_5(2)$ &$K_1(1)\oplus K_3(2)$ &$M_1(1)\oplus M_3(1)\oplus M_4(1)$ \\ \hline \\
			EBRs \\ \hline
			$A_1\uparrow G(3c)$&$\Gamma_1(1)\oplus\Gamma_5(2)$ &$K_1(1)\oplus K_3(2)$ &$M_1(1)\oplus M_3(1)\oplus M_4(1)$ \\
			$B_2\uparrow G(3c)$&$\Gamma_4(1)\oplus\Gamma_6(2)$
			&$K_2(1)\oplus K_3(2)$&$M_1(1)\oplus M_2(1)\oplus M_4(1)$
			\\
			$A_1\uparrow G(1a)$&$\Gamma_1(1)$&$K_1(1)$&$M_1(1)$
			\\ \hline
			\\	
			FTBs \\ \hline
			upper set&$\Gamma_5(2)$&$K_3(2)$&$M_3(1)\oplus M_4(1)$ \\
			lower set&$\Gamma_5(2)$&$K_3(2)$&$M_3(1)\oplus M_4(1)$ 		
		\end{tabular}
	\end{ruledtabular}
\end{table*}


\subsection{Wlison loop spectra}\label{section2b}
Even though the two sets of FTBs both have a zero chern number, their non-trivial topology can be manifested in the Wilson-loop spectrum. First of all, we define the Wilson-loop operator $w_{\boldsymbol{k}}$ \cite{peri2021fragile}, which is constructed by the wave functions ${\psi^1}$ and ${\psi^2}$ of the two FTBs from the same set:
\begin{equation}
w_{\boldsymbol{k}}=\left[{\psi^1}_{\boldsymbol{k}},{\psi^2}_{\boldsymbol{k}}\right],
\end{equation}
where $\boldsymbol{k}$ = $k_1\boldsymbol{b_1}+k_2\boldsymbol{b_2}$, with $k_1$, $k_2$ $\in$ [0,2$\pi$]. This operator is discretized into $N^{2}$ operators when periodic conditions are applied to the boundaries of a N$\times$N TKL. Next, while keeping $k_2$ fixed, we multiply all the discretized Wilson operators along an integral loop in $\boldsymbol{b_1}$ direction \cite{peri2021fragile}:
\begin{equation}
	W(k_2)=\prod\limits_{n=0}^{N-1}w_{\frac{n\cdot2\pi}{N},k_2}^\dagger\cdot w_{\frac{(n+1)\cdot2\pi}{N},k_2}
\end{equation}
By calculating the eigenvalues of $-i\log W(k_2)$, we obtain the Berry phases of the two FTBs along the one-dimensional integral loop that starts at (0,$k_2$). As $k_2$ circulates around the Wilson loop (i.e., a closed path around the BZ along $\boldsymbol{b_2}$ direction), the one-dimensional Berry phases of the two FTBs evolve from -$\pi$ ($\pi$) to $\pi$ (-$\pi$) as shown in Fig.\ref{fig.1c}$\sim$\ref{fig.1d}, contributing opposite winding numbers (n=$\pm$1). This is a $\mathbb{Z}$ classification introduced in the Wilson loop to diagnose the topology of connected subspaces in which every band crosses with each other. In fact, it is the combined symmetry $C_{2z}\mathcal{T}$ that protects these non-trivial windings \cite{bouhon2019wilson,bradlyn2019disconnected,peri2021fragile}. Here are the reasons: (1) After the addition of the NNN hoppings, $\mathcal{T}$ is still preserved and enforces the chern number of each set of FTBs to be zero, which consequently protects the opposite winding numbers. (2) By calculating the $C_{2z}$ eigenvalues of each set of FTBs in BZ, we find that the signs of these eigenvalues are opposite at $\Gamma$ and $M$ (see Table.\ref{table2}). This difference of a minus sign symbolizes a phase difference of $\pi$ between the wavefunctions at the two points, which explains the non-trivial windings of Berry phases from $\Gamma$ to $M$ along the Wilson loop. (3) In this 2D spinless model, $C_{2z}$ rotates $\boldsymbol{k}$ to -$\boldsymbol{k}$ around the out-of-plane $\mathbf{z}$ axis and $\mathcal{T}$ reverses -$\boldsymbol{k}$ back to $\boldsymbol{k}$, keeping every momentum $\boldsymbol{k}$ in BZ fixed. Therefore, $C_{2z}\mathcal{T}$ is preserved everywhere in the BZ, including in the Wilson loop. As a result, the continuity of the evolution of the Berry phases along this loop is ensured by this combined symmetry.

\begin{table}
	\caption{ $C_{2z}$ eigenvalues of the two sets of FTBs at $\Gamma$ and $M$.}
	\label{table2}
	\begin{ruledtabular}
		\begin{tabular}{ccccc}
			
			FTBs&$\Gamma$&$M$\\ \hline
			upper set&$2$ &$-2$
			\\
			lower set&$2$ &$-2$
			\\
	
		\end{tabular}
	\end{ruledtabular}
\end{table}

\subsection{Fractional corner states hosted by an OAL insulator}\label{section2c}

\begin{figure*}
	\centering
	\hspace{0.1in}
	\subfigure[]{
		\includegraphics[width=3in]{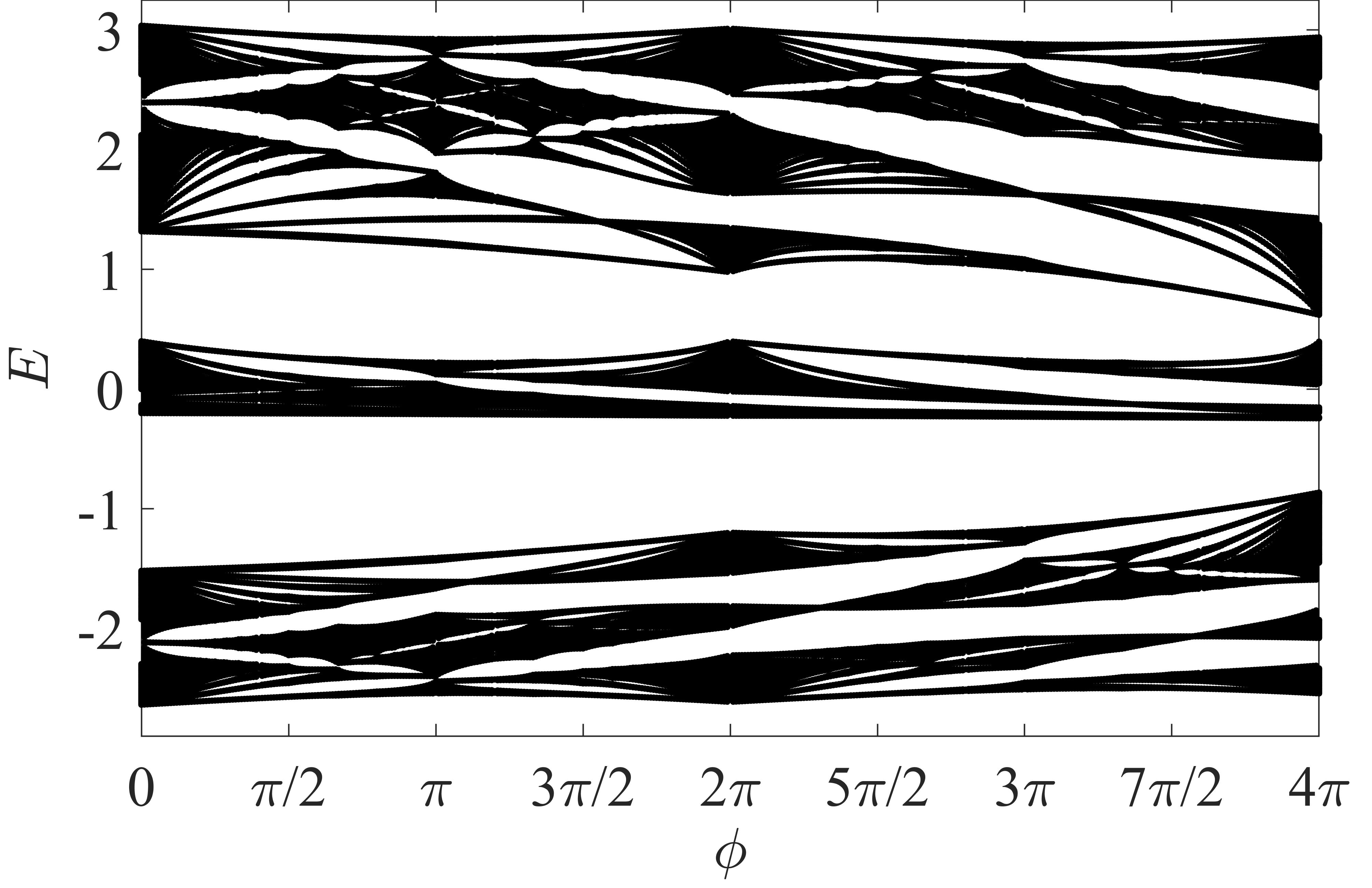}
		\label{fig3a}
	}
	\subfigure[]{
		\includegraphics[width=3in]{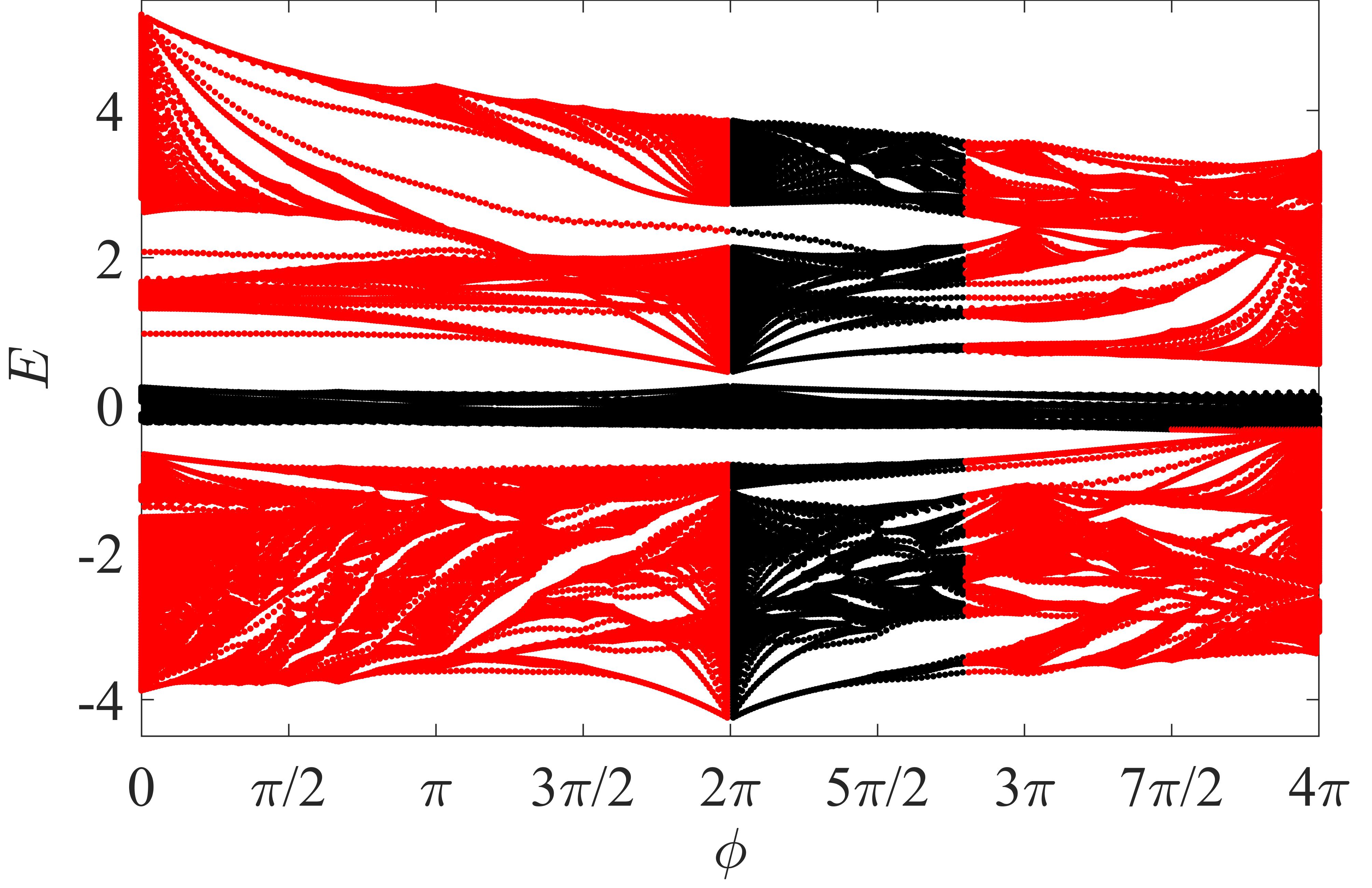}
		\label{fig3b}
	}	
	\subfigure[]{
		\includegraphics[width=3.3in]{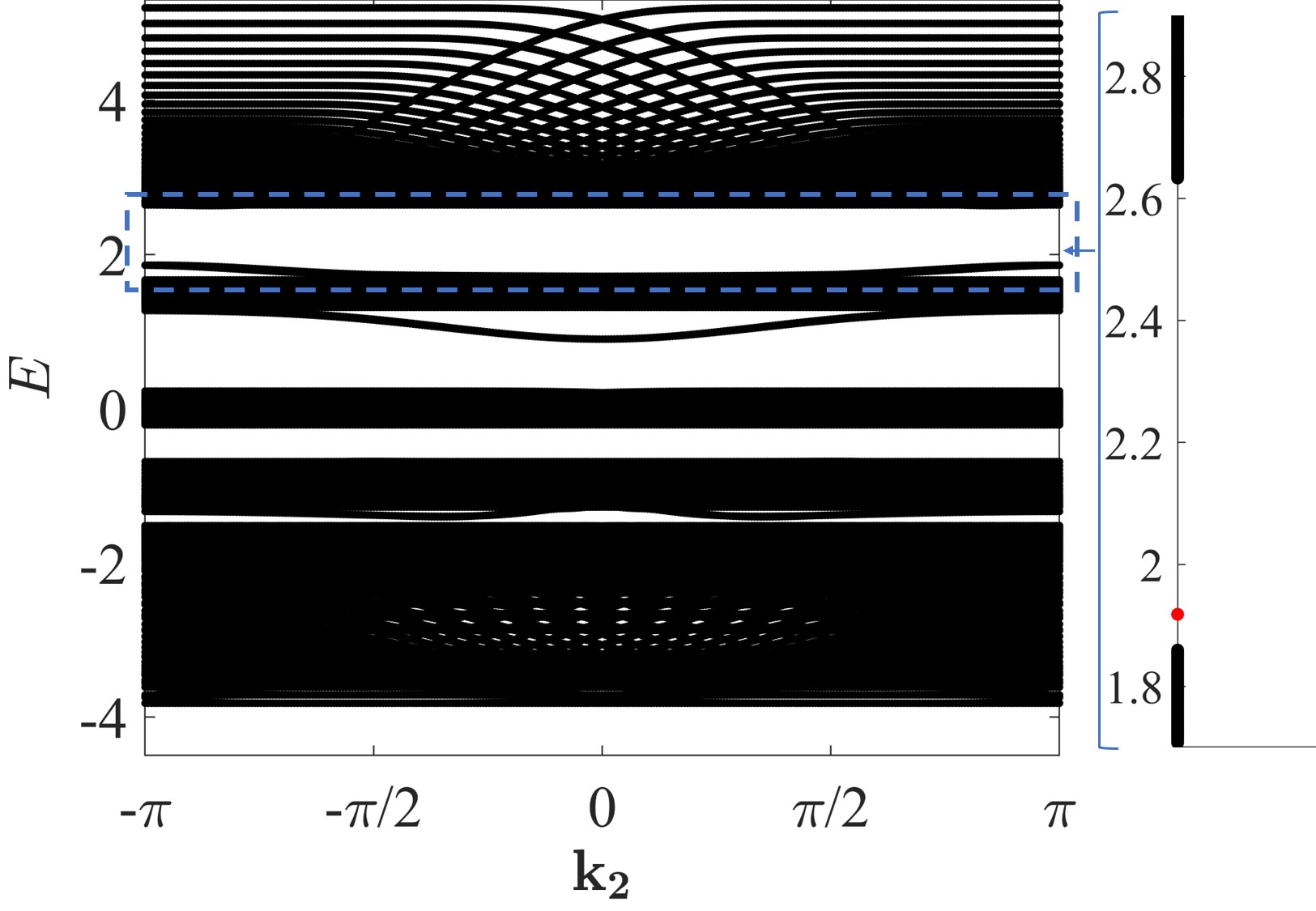}
		\label{fig3c}
	}
	\subfigure[]{
	\includegraphics[width=2.7in]{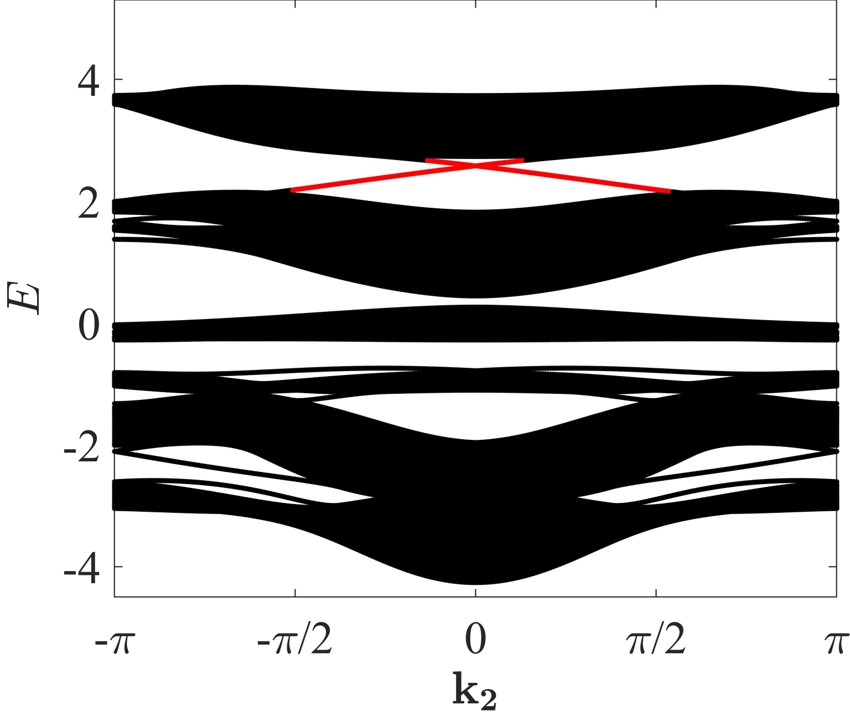}
	\label{fig3d}
}
	
	
	
	\caption{$\left[(a),(b)\right]$ Hofstadter butterfly spectra for (a) $H_{_{\emph{NN}}}^{^{\emph{hof}}}$ and (b) $H_{_{\emph{NNN}}}^{^{\emph{hof}}}$, with the same parameters of Fig.\ref{fig.1b} and Fig.\ref{fig2a}, respectively. 
	In (a), the Hofstadter butterflies of the top and bottom groups expand towards the middle group. In (b), the connected parts of the Hofstadter butterflies are painted red. $\left[(c),(d)\right]$ Energy spectra of $H_{_{\emph{NNN}}}^{^{\emph{hof}}}$ with open boundary condition in $\boldsymbol{a_1}$ direction at (c) $\phi=\frac{2\pi}{q}$($q=200$) and (d) $\phi=2\pi$. In (c), the insert on the right exhibits two degenerate corner states marked by red dots. In (d), chiral edge states are represented by red lines.}
	\label{fig3}

\end{figure*}
We apply the open boundary conditions in $\boldsymbol{a_1}$ and $\boldsymbol{a_2}$ directions to cut the TKL into a 30$\times$30 lattice and then calculate the energy spectrum of $H_{_{\emph{NNN}}}^{^{\emph{TB}}}$ to examine the bulk-boundary correspondence of the FTBs. The results are shown in Fig.\ref{fig2c}, where two isolated degenerate states appear in the gap above the upper set of FTBs. By looking into the distribution of their wavefunctions on the finite TKL, we determine that these states are nothing but fractional corner states, which, according to \cite{luo2021fragile,shang2020second,PhysRevLett.123.186401,lange2021subdimensional,herzog2020hofstadter}, are actually created by the presence of the OAL above them. Given that FTBs are always accompanied by OALs, one might expect this bulk-corner correspondence to be a fingerprint of FTPs. However, we find that an OAL and its corner states appear without the accompany of FTBs, which implies that the presence of OALs is not a sufficient condition to the fragile topology. To demonstrate this phenomenon, first of all, we replace the NNN hoppings with the Haldane-model-like ones [see Fig.\ref{fig.1a}] to break $\mathcal{T}$. We still refer to these terms as the spin-orbit-coupling (SOC) terms even though our model is spinless. They can be considered as the hoppings between orbitals with the same spin (e.g., $\uparrow$ or $\downarrow$). The TB Hamiltonian then becomes
\begin{equation}
	\begin{aligned}
&	 H_{_{\emph{SOC}}}^{^{\emph{TB}}}=H_{_{\emph{NN}}}+H_{_{\emph{SOC}}},\\
&	 H_{_{\emph{SOC}}}=\sum_{\beta^{\prime},\beta}\exp\!\left(i\phi(\boldsymbol{r}_{\beta^{\prime}}-\boldsymbol{r}_{\beta})\right)t_{soc}C_{\beta^{\prime}}^{\dagger}C_{\beta}+H.c. ,
	\end{aligned}
\end{equation}
where $\exp\!\left(i\phi(\boldsymbol{r}_{\beta^{\prime}}-\boldsymbol{r}_{\beta})\right)$=($\pm$)$i$ when the hoppings are anticlockwise (clockwise), as indicated in Fig.\ref{fig.1a}. Apparently, these hoppings are endowed with in-plane chiralities by their phase factors. Hence, $\mathcal{T}$ is broken due to the imaginary phase factors, and all the mirror symmetries of P6mm with mirror planes perpendicular to the plane of Fig.\ref{fig.1a} are also broken, since no in-plane chirality is invariant after undergoing these reflections. Consequently, the space-group symmetry is reduced from P6mm (No.183) to P6 (No.168), which lifts all the degeneracies at $\Gamma$ and $K$ and splits the bottom group into three single bands, as shown in Fig.\ref{fig2d}. The middle band is an OAL represented by the EBR $E_1\uparrow G(1a)$, whereas the other two bands can not be represented by any one-dimensional EBRs or subtractions of EBRs and therefore carry opposite Chern numbers ($C=\pm$1). Naturally, a pair of edge states with opposite chiralities appears in the gaps between the two bands and the OAL, as shown in Fig.\ref{fig2e}. Besides, as shown in Fig.\ref{fig2f}, two pairs of isolated states (each pair consists of two degenerate states) manifest themselves on the corners of the 30 $\times$ 30 TKL, with their energies lying in the gap above the three bands. If all the bands under this gap are filled, this system turns into an obstructed-atomic-limit insulator which hosts fractional corner states. 


\section{Presence of magnetic fields}\label{section3}
\subsection{Hofstadter butterfly spectra}\label{section3a}
To compute the Hofstadter butterfly spectrum, firstly, we choose the Landau gauge $\boldsymbol{A}(\boldsymbol{r})$=(0 , $\phi\boldsymbol{r}\cdot\boldsymbol{a_1}$) to introduce a uniform magnetic field, where $\boldsymbol{A}$ is the magnetic vector potential and $\phi$ is the magnetic flux per unit cell. In this gauge, the $\boldsymbol{r}$-dependence of the nonzero component of $\boldsymbol{A}$ along $\boldsymbol{a_1}$ direction breaks the translational invariance in this direction. However, at a commensurate $\phi$=$2\pi\frac{p}{q}$ (i.e., the two integers $p$ and $q$ are coprime), the broken translational invariance is restored with the cost of making the unit cells q-times larger. Correspondingly, the first BZ shrinks, with $k_1\in$[0,$\frac{2\pi}{q}$] and $k_2\in$[0,2$\pi$]. In the magnetic field, every hopping along a Peierls path acquires an extra phase factor $\exp\!\left(i\theta(\boldsymbol{r}-\boldsymbol{r^{\prime}})\right)$ via the Peierls substitution, where the argument $\theta(\boldsymbol{r}-\boldsymbol{r^{\prime}})$ can be calculated through an integral along the Peierls path \cite{herzog2020hofstadter}:
\begin{equation}
\theta(\boldsymbol{r}-\boldsymbol{r^{\prime}})=\int_{C_{r^{\prime}\rightarrow r}}\boldsymbol{A}(\boldsymbol{r})\cdot d\boldsymbol{r}.
\end{equation}
In this paper, we choose the Peiels paths to be straight lines connecting ionic sites. Now we give the Hofstadter Hamiltonians for the TB models on the TKL:
\begin{equation}
H^{^{\emph{hof}}}=\sum_{i,j,\alpha,\beta}\exp\!\left(i\theta(\boldsymbol{r_{i,\alpha}}-\boldsymbol{r_{j,\beta}})\right)t_{i,j,\alpha,\beta}\cdot C_{i,\alpha}^\dagger C_{j,\beta}+H.c.,
\end{equation}
where $H^{^{\emph{hof}}}$$\equiv$$H^{^{\emph{hof}}}_{_{\emph{NN}}}$ ($H^{^{\emph{hof}}}_{_{\emph{NNN}}}$) for $H_{_{\emph{NN}}}^{^{\emph{TB}}}$($H_{_{\emph{NNN}}}^{^{\emph{TB}}}$). We diagonalize these Hamiltonians in the enlarged magnetic unit cell and show the resulting Hofstadter butterfly spectra in Fig.\ref{fig3a} and \ref{fig3b}. As shown in Fig.\ref{fig3b}, each set of FTBs connects to its complementary OAL when the magnetic field is present, manifesting its non-trivial topology, whereas the middle group represented by the EBR $B_2\uparrow G(3c)$ always stays bounded no matter how strong the magnetic field becomes. Considering that this bounded pattern comes from the trivial nature of the middle group, it is reasonable to expect that the top and bottom groups are also bounded because of their trivial topology captured by the EBR $A_1\uparrow G(3c)$. However, Fig.\ref{fig3a} shows the anomalous expansions of these groups, which becomes an intriguing issue. We conjecture that the $\mathcal{T}$-breaking effect of the magnetic field, which can lift the degeneracies at high-symmetry points and create isolated Chern bands [as in the case of Fig.\ref{fig2d}], causes the expansions. It also explains the connections between FTBs and OALs, since the $\mathcal{T}$-breaking magnetic field can simultaneously break $C_{2z}\mathcal{T}$ and split each set of FTBs into two Chern bands. As for the middle group, it is immune from being affected by the magnetic field because of its ``solid'' trivial nature, i.e., this group of bands adiabatically turns into a trivial flat band when $t^{\prime}$ gradually decreases to zero. Considering that the expanded and bounded groups are represented by different EBRs, we believe that the theory of TQC may catch some implicit differences between these two trivial band structures, which are waiting for more explorations.

\subsection{Topologcial phase transition from a SOTP to a FOTP}\label{section3b}
The expanded and connected patterns of Fig.\ref{fig3a} and \ref{fig3b} indicate that non-trivial LLs can be shifted under the action of a magnetic field with increasing strength, which enables a topological phase transition from a SOTP to a FOTP in the TKL. Firstly, we search for a gap wherein corner states already exist before the presence of a magnetic field. When all the levels under this gap are filled, a second-order topological phase appears on the lattice. After the shifts of the LLs with non-zero Chern numbers from the unoccupied subspace to the occupied subspace across the gap, a first-order topological phase appears and exhibits topological states on edges instead of on corners, which symbolizes the completion of the transition. On the basis of the discussion in Sec.\ref{section2c}, we choose the gap between the upper set of FTBs and its complementary OAL as our candidate. While the magnetic field is barely introduced ($\phi$=$\frac{2\pi}{q}$), the second-order topological phase is preserved since the corner states still exist in this gap [see Fig.\ref{fig3c}]. At $\phi$=$2\pi$, the Chern number of the occupied bands becomes 1 from zero and the corner states are submerged into chiral edge states [see Fig.\ref{fig3d}], which indicate the transition to the first-order topological phase.
Generally speaking, it is difficult but not impossible to create such a strong magnetic field that squeezes a magnetic flux quantum into one unit cell. Recently, researchers have realized some many-body phases of twisted-bilayer graphene at 2$\pi$ flux in experiments \cite{herzog2021reentrant,das2021observation}, encouraging us to consider realizing this transition in a twisted-bilayer TKL.        


\section{Conclusion and discussion}\label{section4}

Considering the recent popularity of kagome metals \cite{yu2021concurrence,zhang2021pressure,ortiz2020cs} and insulators with non-trivial topology, we believe that it is of great interest to explore the topological physics behind the more frustrated kagome-like lattices, which can simultaneously enhance Wannier obstructions and the electronic correlations. Our work is a very beginning study in this direction. In this work, we diagnose the fragile topology of the two band structures with different energy spans. The windings of Berry phases and the connected Hofstadter butterflies confirm their non-trivial topology. During the calculations, we realize that magnetic fields can release implicit details of topological band structures to create and shift non-trivial LLs, which eventually gives rise to the connected Hofstadter spectrum. However, a trivial band structure represented by $A_1\uparrow G(3c)$ also exhibits an expanded pattern in the magnetic field, which is different from the bounded group represented by $B_2\uparrow G(3c)$. We believe that the trivial nature of the expanded group is less ``solid'' than the bounded group and that the difference between these two trivial band structures is captured by the theory of TQC. Perhaps there is another classification between the trivial and fragile topology. By examining the bulk-boundary correspondence, we find that the presence of OALs and corner states is not a sufficient condition to fragile topology, which makes us wonder whether it is possible to find a FTP without the presence of complementary OALs. It is of great interest to explore the bulk-boundary correspondence of this kind of FTPs. Last but not least, considering that it is experimentally realizable to reach one quantum flux in the moire unit cell of twisted-bilayer structures, we are inspired to explore possible applications of the topological phase transition in a twisted-bilayer TKL.


\section{ACKNOWLEDGMENTS}
We thank Zhong-Bo Yan for enlightening advices, and Jun Li, Jia-Zheng Ma, Zhi-Hui Luo, Shan-Bo Zhou, Biao Lv, Xun-Wu Hu, Ze-Nan Liu for helpful discussions. This project is supported by NKRDPC-2017YFA0206203, NKRDPC-2018YFA0306001, NSFC-11974432, NSFC-92165204, GBABRF-2019A1515011337, Shenzhen Institute for Quantum Science and Engineering (Grant No. SIQSE202102), and Leading Talent Program of Guangdong Special Projects (201626003).

\nocite{*}

\bibliography{paper_reference}

\end{document}